\documentclass[twocolumn,prl,aps,superbib,tightenlines,floatfix]{revtex4}
\usepackage{amsfonts,amsmath,amssymb,amsthm,bm}
\usepackage{graphicx}
\begin{document}
\bibliographystyle{apsrev}
\title{Effects of Isotope Substitution on Local Heating and Inelastic Current in Hydrogen Molecular Junctions}
\author{Yu-Chang Chen}
\email{yuchangchen@mail.nctu.edu.tw}
\affiliation{Department of Electrophysics, National Chiao Tung University, 1001 Ta Hsueh
Road, Hsinchu, Taiwan 30010 }
\begin{abstract}

Using first principle approaches, we investigate the inelastic features in the hydrogen molecular junction.  We observe that local heating and
inelastic current have significant isotope-substitution effects. The junction instability is also relevant to the isotope substitution.
We predict that the HD junction has the smallest breakdown voltage compared with the H$_{2}$ and D$_{2}$ junction in the optimized geometry.
The selection rule for modes that significantly contribute to the inelastic effects is related to the component of vibration along the direction of electron transport.

\end{abstract}
\maketitle

Building electronic circuits from molecules is an inspiring idea. \cite{Aviram}
In the past decade, motivated by this revolutionary idea, intensive theoretical
and experimental studies have been conducted to explore the charge transport
in a single molecule junction. Single molecule junction
is the basic building block for electronic components in molecular electronics.
Understanding the transport properties of the basic building block is of key
importance in the design of a new form of electronic circuit at the atomic and
molecular levels. \cite{Reed,Tao,Ratner2,Galperin}.

\begin{figure}
  \includegraphics[width=8cm]{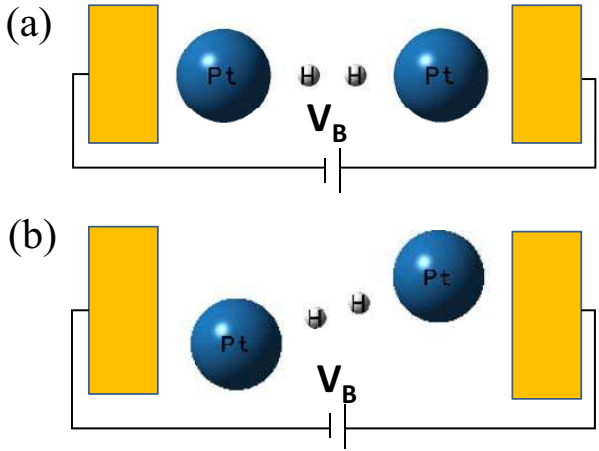}\\
  \caption{(color online) Schematic diagrams for the systems investigated: (a) the optimized geometry of the H$_{2}$ junction. The Pt-H-H-Pt atoms are sandwiched between two bulk Pt electrodes kept at a certain external bias with the chemical potentials of the left/right electrodes are shifted, i.e., the source drain bias is V$_{SD}$=(E$_{FR}$-E$_{FL}$)/e. The Pt electrodes are described as Jellium model with the interior electron density of Pt electrodes is taken close to the value of metallic Pt ($r_{s}\approx3$). The H-H, Pt-H, and Pt-Jellium distance are 1.615, 4.051, and 2.0 a.u., respectively; (b) a specific geometry of the H$_{2}$ junction. The hydrogen molecule rotates an angle of $15^{\circ}$ due to asymmetric contacts.}
\label{fig1}
\end{figure}

The hydrogen junction is an ideal testbed to compare theories and experiments.
Despite the simplicity of the hydrogen junction, the data reported for
the conductance varying from $0.2$ to $1.0~$G$_{0}$ have provoked
controversy. \cite{Smit,Cuevas,Carcia,Csonka,Qi,Thygesen}
The problem of inelastic electron-vibration interactions remains unsettled.
For example, the features of current-voltage characteristics
at a bias around $65~$mV are attributed to the electron-vibration
interactions. \cite{Smit,Djukic} However, a theoretical group has
reported calculations to support the idea that the static atomic
structure can still cause a pronounced resonance at this bias. \cite{Qi}
The mode identification and the selection rule for modes that significantly contributed to
the inelastic electron tunneling spectroscopy (IETS) have not reached unanimity.
The identification of the spike in $dI/dV$ at the Pt-H$_{2}$ system as the transverse
vibration mode is still controversial. \cite{Djukic2}

To clarify the fundamental inelastic electron transport properties, we performed
comparative studies on inelastic profiles among the H$_{2}$, HD, and D$_{2}$ junction.
The simplicity of the hydrogen junction allows the inelastic profiles to provide clear
information for understanding how the vibration modes contribute to inelastic effects
in the nano scale junction. Owing to the small mass of the hydrogen atom, isotope
substitution has a great influence on inelastic effects. While the isotope substitution does not affect
the contact geometry and the bonding strength, it can serve as a controllable
factor that is sensitive enough to explore the fundamental properties of inelastic effects
in the nano scale junction. Moreover, the vibrations excited before the
breakdown voltage of the hydrogen junction are acoustic-like modes connected to
contact geometry and bonding strength. This is in sharp contrast to the IETS and HREELS in
the alkyle junction where all observed modes in experiments are
optical-like modes more relevant to internal vibrations of the
molecule. \cite{Wang,Ratner3,Yluo,Kushmerick,Brandbyge,Dicarlo,alkane} As
such, inelastic profiles in the hydrogen junction can provide more information
on contact geometry and bonding strength in the molecular junctions. Although heating
is an important effect relevant to the stability and performance of device, very few
attempts have been made that addressed the nano junction. \cite{Huang}

\begin{figure}
  \includegraphics[width=8cm]{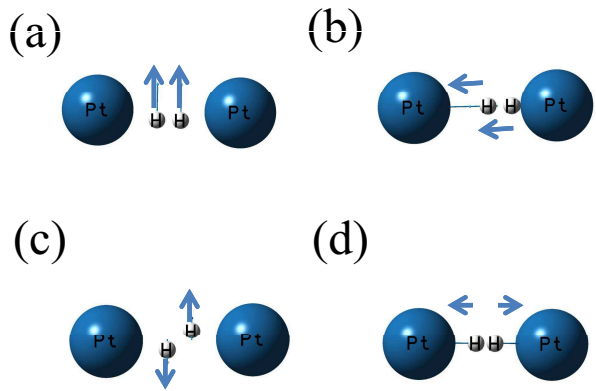}\\
  \caption{(color online) Schematic diagrams for the vibrational modes of the H$_{2}$ molecule. Six vibrational modes are shown in (a) for the modes 1 and 2 associated with a rigid motion of the H$_{2}$ molecule perpendicular to the line connecting two Pt atoms; (b) for the mode 3 associated with a rigid motion of the H$_{2}$ molecule along the line connecting Pt atoms; (c) for the modes 4 and 5 associated with a transverse motions of the H atoms; (d) for the mode 6 associated with an internal vibration of the H$_{2}$ along the line connecting Pt atoms. The energies and vibration characteristics of these modes are shown in Table I for the linear junction with a geometry depicted in Fig.~\ref{fig1}(a).}\label{fig2}
\end{figure}

In our results, we observed interesting inelastic features in the
hydrogen junction. In the optimized geometry (Fig.~\ref{fig1}(a)),
The electron-vibration interaction is suppressed by the mass
of the molecule. The electron-vibration interaction is found to be enhanced by
asymmetric molecule, which leads to the most prominent
heating and inelastic correction to the current in the HD junction. Since the
current-induced force, irrelevant to the isotope substitution,
is found to be small, we predict that the HD junction is
the most unstable owing to local heating as the major breakdown mechanism.
This feature can be verified by experiments.
In a possible rotated geometry as shown in Fig.~\ref{fig1}(b), we observe that
all modes significantly contribute to inelastic profiles. It further confirms
that the selection rule of important modes in the inelastic profiles is
connected to the component of molecule vibration along the direction of charge transport.
Thus, inelastic profile can a sensitive tool to probe the contact geometry and bonding strength
in the nano scale junction.

\begin{table}
\caption{Vibration Modes  depicted in Fig.~\ref{fig2}}
\begin{tabular}{ccccc}
\hline\hline
modes & $\omega _{H_{2}}$ & $\omega _{HD}$ ($\frac{\omega _{H_{2}}}{\omega
_{HD}}$) & $\omega _{D_{2}}$ ($\frac{\omega _{H_{2}}}{\omega _{D_{2}}}$) &
features \\ \hline\hline
1 & 36.8 & 30.0 (1.22) & 26.1 (1.41) & transverse \\
2 & 36.8 & 30.0 (1.22) & 26.1 (1.41) & transverse \\
3 & 64.8 & 52.9 (1.22) & 45.9 (1.41) & longitudinal \\
4 & 115.9 & 101.1 (1.15) & 82.1 (1.41) & transverse \\
5 & 115.9 & 101.1 (1.15) & 82.1 (1.41) & transverse \\
6 & 314.9 & 273.4 (1.15) & 222.7 (1.41) & longitudinal \\ \hline
\end{tabular}

\footnotetext{Unit of energies in meV}
\label{table1}
\end{table}

We will begin our discussion by considering the inelastic current and
heating effects in the linear hydrogen junction (Fig.~\ref{fig1}(a)).
We study these effects by the first order perturbation theory based on the second
quantized formalism where the wave functions are obtained from density functional
theory (DFT). \cite{chen1} The scattering wave functions of the whole
system are calculated by solving the Lippmann-Schwinger equation iteratively
until the self-consistency is obtained. \cite{Lang} The exchange and
correlation energy are calculated at the level of the local density
approximation. \cite{limit} These wave functions are applied to calculate current and heating.
The heating phenomenon in nanojunctions has been measured experimentally
in the alkane junctions in C-AFM, and the data are in reasonable agreement with
our theoretical calculations. \cite{Huang,Huang2}

At first glance, the conductance of the hydrogen molecule is expected to be
small because of a closed shell configuration. However, the hybridization of states
between hydrogen and Pt atoms can create new channel for current. \cite{Thygesen} As
a result, the magnitude of conductance, irrelevant to the isotope
substitution, can be quite large around $1.0~$G$_{0}$in the hydrogen junction.
This feature is verified by the analysis of partial DOS. In our
results, the current-voltage characteristic in the Pt-H-H-Pt junction is quite
linear, with a conductance around $G\approx0.9~$G$_{0}$. We find that the
S-wave state, mostly contributed by the hydrogen atoms, has negligible
contribution to the conducting state. This feature supports the argument that the
origin of charge transport comes mainly from the H-Pt hybridized states.
\cite{Thygesen}

We briefly introduce the theory of the electron-vibration
interactions. By applying the field operator with the wave functions obtained
from the static DFT, the electron-vibration interaction has a second
quantized form, \cite{chen1}

\begin{eqnarray}
H_{el-vib} &=&\sum_{\alpha ,\beta ,E_{1},E_{2},j,\nu }\left( \sum_{i,\mu }%
\sqrt{\frac{\hbar }{2M_{i}\omega _{j\nu }}}A_{i\mu ,j\nu
}J_{E_{1},E_{2}}^{i\mu ,\alpha \beta }\right)   \notag \\
&&\cdot a_{E_{1}}^{\alpha \dag }a_{E_{2}}^{\beta }(b_{j\nu }+b_{j\nu }^{\dag
}),  \label{elph}%
\end{eqnarray}
where $M_{i}$ is the mass of the $i-th$ atom, $A_{i\mu,j\nu}$ is a canonical
transformation between normal coordinates and Cartesian coordinates
satisfying $%
{\displaystyle\sum\nolimits_{i,\mu}}
A_{i\mu,j\nu}A_{i\mu,j^{^{\prime}}\nu^{^{\prime}}}=\delta_{j,j^{^{\prime}}%
}\delta_{\nu,\nu^{^{\prime}}}$; $J_{E_{1},E_{2}}^{i\mu,\alpha\beta}$ is the
coupling constant related to the electron and vibration of atoms, which can
be calculated from the pseudopotential of atoms
and wavefunctions; $a_{E}^{(L,R)}$ is the annihilation operator
for an electron with energy E incident from the left or right electrode, respectively;
$\hbar\omega_{j\nu}$ represents the energy of the $j\nu$-th normal mode and $b_{j\nu}$
is the annihilation operator corresponding to that mode. The strength of
electron-vibration coupling is affected by the isotope substitution related to the mass
of the molecule shown in Eq.~(\ref{elph}).

As the first step in our analysis, we focus on inelastic
profiles in an optimized linear geometry as shown in Fig.~\ref{fig1}(a).
Two heavy Pt electrodes, which are inert owing to heavy masses, are simulated as
two Pt atoms with infinite mass for simplicity. The geometry is relaxed
in gauusian03 with the total energy of the system around $-380.15$ Hartree.
The geometry is stable and requires $0.53 eV$ to take the H$_{2}$ molecule
away from the junction. In Table~\ref{table1}, we list the energies of all the modes, while the
schematics of vibrations for the H$_{2}$ molecule are shown in Fig.~\ref{fig2}.
Mode 3 is the one responsible for significant inelastic current and heating
because of its longitudinal nature. Mode 6 is associated with a longitudinal internal
vibration. The energy of this mode is too large to be excited before the breakdown
bias around $200~$mV found in experiments. \cite{Smit,chen3} The rest
of the modes have insignificant inelastic effects owing to their transverse natures.

\begin{figure}
\includegraphics[width=8cm]{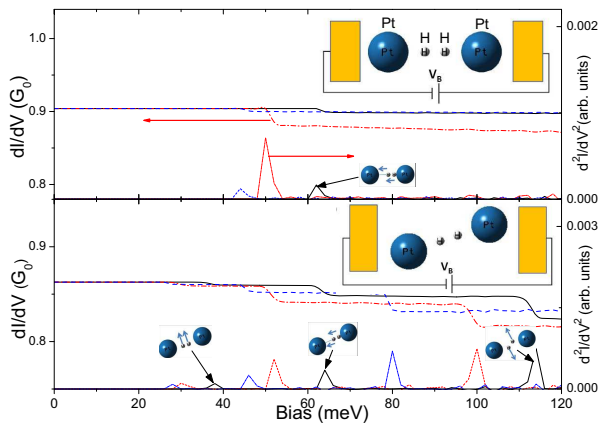}
\caption{(color online) Differential conductance (left axis) and absolute value of $d^{2}I/dV^{2}$ (right axis) as a function of bias for the H$_{2}$ (black solid line), HD (red dotted-dashed line), and D$_{2}$ (blue dashed line) molecule sandwiched between the optimized linear junction (upper panel) and the junction of with asymmetric contacts(lower panel). The schematics show the noticeable modes contributed to the inelastic current in the H$_{2}$ junction.}
\label{fig3}
\end{figure}

In the linear hydrogen junction, the position of the discontinuities in the differential
conductance is affected by the isotope substitution as shown in the upper panel of Fig.~\ref{fig3}.
In the H$_{2}$ junction, an abrupt change in the differential conductance
appears at a bias around $65~$mV in the $d^{2}I/dV^{2}$ spectrum. This peak is
associated with mode 3 which has longitudinal character. This mode is associated
with motion of the molecule as a rigid body. Accordingly, the energies are perfectly scaled
by $\omega _{H_{2}}/\omega_{HD}\varpropto \sqrt{m_{HD}/m_{H_{2}}}\approx 1.22$. The rest
modes with transverse characters are too weak to be resolved.

The magnitude of electron-vibration interaction is also affected by
the isotope substitution. Let us compare the magnitude of discontinuities
in the differential conductance in the linear H$_{2}$ and D$_{2}$ junctions.
These two junctions have the same mechanical structures except for different masses.
The electron-vibration coupling is related to the mass of ions by $1/\sqrt{M_{i}}$.
Thus, the H$_{2}$ junction has a larger discontinuity
in the differential conductance because of smaller mass. In the HD junction, the
electron-vibration interaction is enhanced by asymmetric mass distribution.
As a result, the HD junction has the largest change in the differential conductance.

Local temperatures in the linear H$_{2}$, HD, and D$_{2}$ junctions are further
investigated to elucidate the role of isotope substitution. Two major processes
lead to an equilibrium local temperature in a nanojunction. One is due to the
electron-vibration interaction that occurs in the atomic region
of the junction; the other is due to the dissipation of heat energy to the bulk
electrodes via thermal transport. \cite{Geller} We assume that the energy generated
in the atomic region via inelastic electron-vibration scattering and the
energy dissipated to the electrodes via thermal current finally reach equilibrium,
such that a well-defined local temperature can be calculated in the atomic region.
We evaluate the probability of the energy exchange via electron-vibration
interactions by the Fermi golden rule for the four first-order scattering
mechanisms. These mechanisms correspond to electrons incident from the right
or left electrode that relax (cool) or excite (heat) the energy level of the
normal mode vibrations in the atomic region. The total thermal power generated
in the junction can be calculated as the sum of all vibrational modes of
the above four scattering mechanisms. Detailed theories can be found in Ref. \cite{chen1}.

\begin{figure}
\includegraphics[width=8cm]{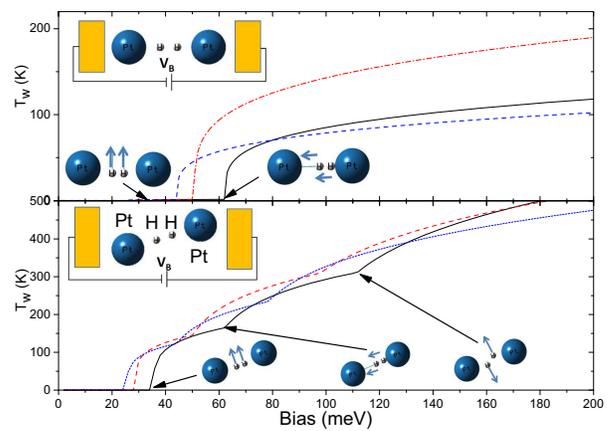}
\caption{(color online) Local temperature as a function of bias for the H$_{2}$ (black solid line),
HD (red dotted-dashed line), and D$_{2}$ (blue dashed line) molecule sandwiched between the optimized linear junction (upper panel) and the junction with asymmetric contacts(lower panel). The schematics show the noticeable modes contributed to heating in the H$_{2}$ junction.}
\label{fig4}
\end{figure}

Heating induced by the electron-electron interaction is neglected in present work since the
contribution might be small compared with the electron-vibration
interaction at small biases. \cite{Diventra,Huang} In the upper panel of Fig.~\ref{fig4}
we plot the equilibrium temperature of
the molecule as a function of applied voltage in the linear H$_{2}$, HD,
and D$_{2}$ junctions. Weak heating phenomena occur
at small biases corresponding to modes 1 and 2. These modes have very small contribution
to heating due to their transverse vibration characters. Major heating profiles are
observed at onset biases around $64.8~$mV, $52.8~$mV, and $45.9~$mV for
the H$_{2}$, HD and D$_{2}$ junction, respectively. Those heating
profiles are related to mode 3 with longitudinal vibration character.

The interplay between the onset bias and the strength of electron-vibration
interaction caused by the isotope substitution leads to interesting heating profiles.
For symmetric molecules in the linear junction (Fig.~\ref{fig1}(a)), the magnitude
of electron-vibration coupling is enhanced by the smaller mass (the H$_{2}$ junction)
which leads to a larger
heating rate, whereas the molecule with a larger mass (the D$_{2}$ junction) has
a smaller onset bias. Consequently, a crossing in the equilibrium temperatures
for the H$_{2}$ and D$_{2}$ junctions occur at $80~$mV as shown in the upper panel
of Fig.~\ref{fig4}. For the biases smaller than $80~$mV,
the local temperature in the D$_{2}$ junction is larger due to a smaller onset bias.
For the biases larger than $80~$mV, the local temperature in the H$_{2}$ junction
is larger due to the larger electron-vibration coupling. For the HD junction,
the asymmetric mass distribution leads to the enhancement of electron-vibration
interaction. It accordingly induces the largest heating rate after
the longitudinal normal mode is excited.

The junction stability is also relevant to isotope substitution. In
atomic-sized Al wires, the current-induced force is the major mechanism
that causes mechanical breaking, whereas in alkane junctions heating is
the dominant breaking mechanism. \cite{chen1,Yang} In our results, the
current-induced forces on the hydrogen atoms have a small value of $0.38~$nF
at $200~$mV. We thus expect that the major breakdown mechanism will be caused by
local heating rather than the current-induced force. Because local heating
is susceptible to effects of isotope substitution in contrast to the
current-induced force where the isotope substitution plays no role,
in the optimized geometry (Fig.~\ref{fig1}(a)), we predict that compared with
the H$_{2}$ and D$_{2}$ junctions, the HD junction has the smallest breakdown
voltage caused by the most prominent heating.

As the detailed configurations in the contact region are unknown
in the real experiment, the optimization of contact is not our present
concern. We  will concentrate on how the orientation of the
molecule affects the inelastic profiles. We may now proceed to consider a specific
configuration of the hydrogen junction where the hydrogen molecule rotated
at an angle of $15^{\circ}$ as shown in Fig.~\ref{fig1}(b). The modes
with transverse characters in the linear junction turn out to have large component
of vibration along the direction of current in rotated geometry. The responses
of inelastic current and local heating to the biases are shown in the
lower panels of Figs.~\ref{fig3} and \ref{fig4}, respectively.
In contrast to the optimized case (Fig.~\ref{fig1}(a)) where only mode 3
had significant contribution to inelastic effects, all modes that can
be excited cause prominent features in the inelastic profiles.
These features further illustrate the selection rule for modes that
significantly contribute to the inelastic effects related to the
component of vibration along the direction of electron transport.

In conclusion, we have calculated various transport properties in H$_{2}$, HD,
and D$_{2}$ molecular junctions in linear and rotated geometries.
In the linear geometry, the I-V characteristics is linear with
conductance $G\approx0.9~$G$_{0}$. We investigate
how the isotope substitution can influence the inelastic current and
local heating. The interplay between vibration energies and the
electron-vibration coupling strength leads to interesting heating profile.
We predict that the HD junction has the smallest breakdown voltage due to
the enhancement of electron-vibration interaction. This feature is
measurable and could give an insight into the mechanisms responsible for
the system breakdown. We also investigate the inelastic effects in rotated
geometry. The inelastic features further illustrate the selection rule
for modes that significantly contribute to the inelastic effects related to the
component of vibration along the direction of electron transport.

I am grateful to Prof. M. Di Ventra, Dr. M. Chshiev and Prof. C. S. Chu for helpful
discussions. This work is supported by Taiwan NSC 96-2112-M-009-037, MOE ATU, NCTS,
and computing resources from NCHC.


\begin{thebibliography}{99}
\bibitem {limit}Density functional theory in local density approximation may underestimate the HOMO-LUMO gap for certain molecules. Quantitative description may not be accurate for insulating molecule. However, the hydrogen junction is in conducting state which lies in the reliable regime that the current approach can be applied.
\bibitem {chen3}We employed the Gaussian 03 program to evaluate the vibrational
modes of H$_{2}$, D$_{2}$, and HD molecule bonding to two Pt atoms at DFT
level.
\end{thebibliography}
\end{document}